# Hazes and Clouds in a Singular Triple Vortex in Saturn's Atmosphere from HST/WFC3 multispectral imaging

J.F. Sanz-Requena[a,b], S. Pérez-Hoyos[c], A. Sánchez-Lavega [c], T. del Rio-Gaztelurrutia [c], Patrick G.J. Irwin[d]

a) *Departamento de Ciencias experimentales. Universidad Europea Miguel de Cervantes, Valladolid, Spain.*
b) *Departamento de Física Teórica, Atómica y Optica. Universidad de Valladolid, Spain*
c) *Departamento de Física Aplicada I, Escuela Técnica Superior de Ingeniería, Universidad del País Vasco, Bilbao, Spain*
d) *Atmospheric, Oceanic and Planetary Physics, University of Oxford, Clarendon Laboratory, Parks Road, Oxford OX1 3PU, UK*

**Abstract**

In this paper we present a study of the vertical haze and cloud structure over a triple vortex in Saturn's atmosphere in the planetographic latitude range 55ºN-69ºN (del Rio-Gaztelurrutia et al. , 2018) using HST/WFC3 multispectral imaging. The observations were taken during 29-30 June and 1 July 2015 at ten different filters covering spectral range from the 225 nm to 937 nm, including the deep methane band at 889 nm. Absolute reflectivity measurements of this region at all wavelengths and under a number of illumination and observation geometries are fitted with the values produced by a radiative transfer model. Most of the reflectivity variations in this wavelength range can be attributed to changes in the tropospheric haze. The anticyclones are optically thicker ($\tau \sim 25$ vs $\sim 10$), more vertically extended (~ 3 gas scale heights vs ~ 2) and their bases are located deeper in the atmosphere (550 mbar vs 500 mbar) than the cyclone.

**Keywords:**
Saturn, Saturn Atmospheres, Radiative Transfer, Hubble Space Telescope Observations.

**Highlights**

- We present the cloud and haze structure in a triple vortex in Saturn´s atmosphere.

- Most of the reflectivity changes are related to the properties of the tropospheric haze.
- In the anticyclonic region we find a higher particle number density than in the cyclonic region
- The aerosol optical thickness and vertical extent agree with upwelling in the anticyclones.

**1. Introduction**

The zonal wind profile of Saturn's upper clouds is approximately symmetrical, with a strong prograde equatorial jet and four other eastward jets in the northern and southern hemispheres (Sánchez-Lavega et al., 2000; García-Melendo et al., 2011). The jet at 65ºN planetographic latitude (PG) (in this paper, all latitudes are given in planetographic units, except stated otherwise) has a singular structure, with a double peak (del Rio-Gaztelurrutia et al., 2018, del Genio et al., 2009) that marks two different dynamical regions that are very close in latitude. Both have a similarly high eastward velocity and the ambient vorticity facilitates the coupling of opposite voriticity ovals located to the north and south of the velocity local minimum, as shown in Figure 1 (del Rio-Gaztelurrutia et al., 2018). This double jet seems to be permanent having been observed since Voyager times (Sánchez-Lavega et al., 2000; García-Melendo et al., 2011).

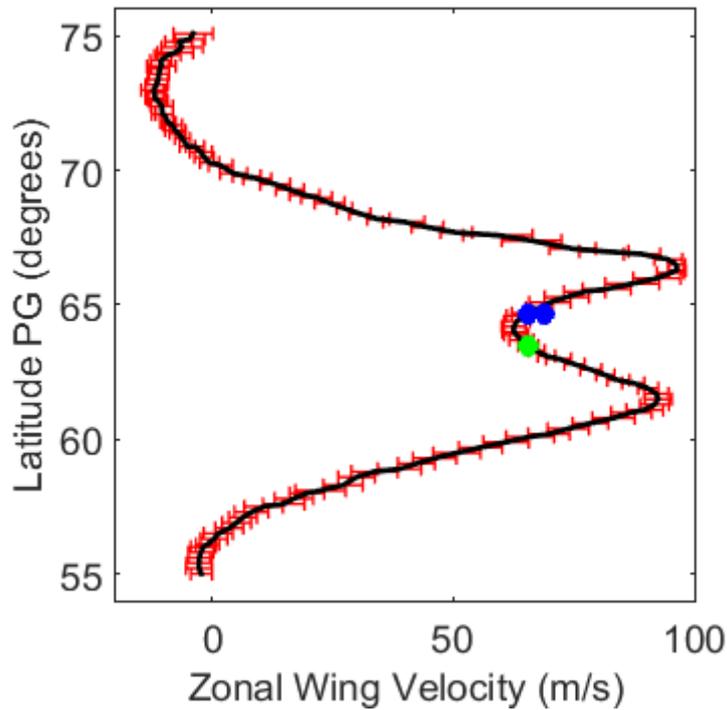

Figure 1: Zonal wind profile for the region of interest (del Río-Gaztelurrutia et al., 2018). The blue points indicate the location of the anticyclones and the green point indicates the location of the cyclone.

In fact, at the latitude of this double peak, a system of three vortices, a cyclone and two anticyclones can be tracked in Cassini ISS images since the beginning of 2012 (del Rio-Gaztelurrutia et al., 2018), confirming that vortices in Saturn can be long-lived (Trammel et al., 2016; del Rio-Gaztelurrutia et al., 2010). We shall refer to the triple vortex system as Anticyclone-Cyclone-Anticyclone abbreviated as the ACA system.

In Saturn, the detection of vortices using ground-based telescopes used to be complicated (del Rio-Gaztelurrutia et al., 2018), and most of our knowledge of these systems comes from space-based observations. Vortices were first detected during the Voyager flybys in 1980–81 (Smith et al., 1981, 1982; Ingersoll et al., 1984; García-Melendo et al., 2007), and then by the Hubble Space Telescope (HST) (Sánchez-Lavega et al., 2004) and the Cassini spacecraft (Vasavada et al., 2006; Trammel et al, 2016; Ingersoll et al., 2018; Sayanagi et al., 2019; Sánchez-Lavega et al., 2019). In more recent times, the improvement of observation techniques has allowed the observation of vortices from Earth even with small sized telescopes and in May 2015, amateur observers detected a disturbance that started at the location of the triple vortex system, which had been previously observed in their images as a dark spot. The perturbation evolved fast, extending rapidly in longitude. The orbits of the Cassini spacecraft at the time were not favorable for the observation of the region, and so we were granted Director Discretionary Time at the Hubble Space Telescope (HST) to observe the region before the perturbation faded away (del Rio-Gaztelurrutia et al., 2018). More recently, in 2018, a convective outbreak occurred in the cyclonic side of the poleward jet disturbing the latitude band from ~ 65°N to 76°N (Sánchez-Lavega et al., 2019).

The study of the existence of long-lived vortices and their evolution is an excellent way to increase our understanding of the atmospheric conditions below the observable upper clouds (García-Melendo et al., 2007; del Río-Gaztelurrutia et al., 2010). An essential factor in understanding the atmospheric dynamics of vortices is the knowledge of the vertical distribution of the haze and clouds used as tracers, and to achieve this knowledge we need to determine also the physical and optical properties of the haze particles and clouds in Saturn's stratosphere and upper troposphere (Sanz-Requena et al., 2018). Our current understanding of Saturn's clouds and hazes is constrained by several decades of remote sensing data (e.g., Pérez-Hoyos et al., 2005; Karkoschka 2005; West et al., 2009) and a usual model is to consider a three-layered aerosol structure formed by a thin stratospheric haze and a denser tropospheric haze, both above a thick cloud layer (Roman et al., 2013).

The goal of this paper is to evaluate the structure of the clouds and hazes and the distribution of aerosols and particles and their properties in the upper troposphere and lower stratosphere in the region of the triple vortex and its environs area, using HST/WFC3 multispectral imaging.

The paper is organized as follows: Section 2 is devoted to a short description of the observations used in this work. Section 3 covers the radiative transfer model, including a description of the vertical cloud structure model and its a priori assumptions. Results are presented and discussed in Section 4, including an analysis of the sensitivity to the model parameters. Results are discussed in Section 5 in terms of the local dynamics and a summary of the main conclusions of this work is presented in Section 6.

**2. Observations**

*2.1. Description of the observations*

In this study, we have used 42 images taken with the Wide Field Camera 3 (WFC3) onboard HST. The images were taken with a variety of filters in three different orbits, on June 29–30 and July 1, 2015. We show in Figure 2 a representative set of these images.

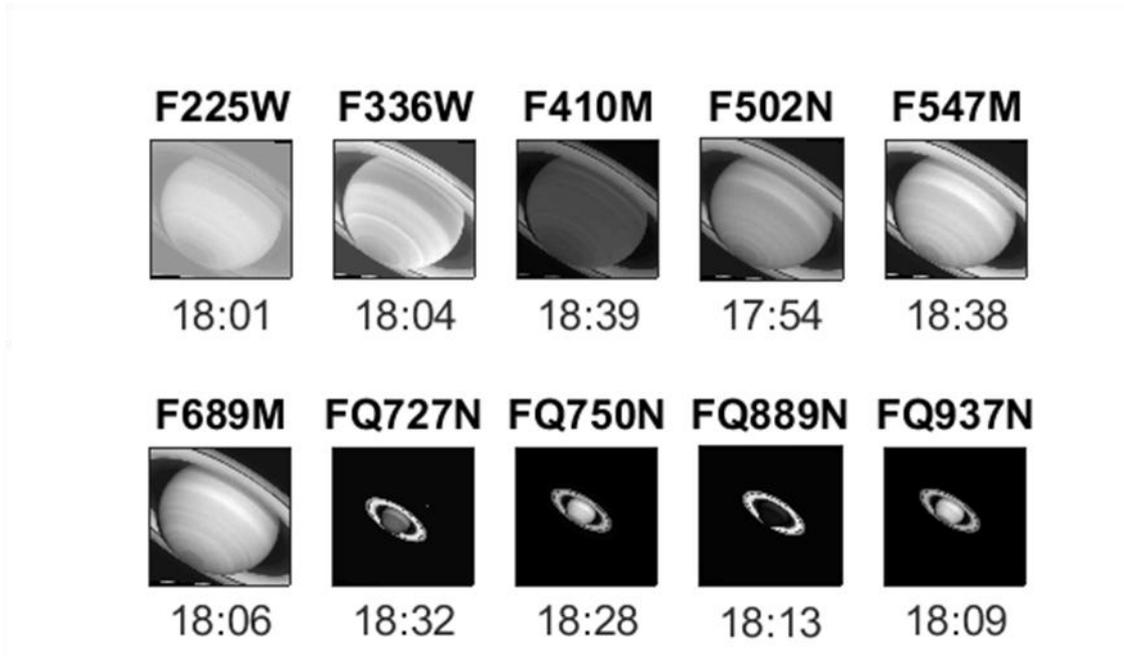

Figure 2: Images taken on July 1, 2015 for the ten filters used in this work. Note that quad filters (FQ727N, FQ750N, FQ889N and FQ937N) are binned for a better signal-to-noise ratio.

Table 1 summarizes the observations used in this work and some of the geometric parameters that characterize them. All images have been photometrically calibrated (Dressel, 2019), and navigated and cylindrically projected with the LAIA software, developed by J.A. Cano (Grup d'Estudis Astronomics, GEA) (Sanz-Requena et al., 2012 and Pérez-Hoyos et al., 2005). We assumed a 10% error in absolute calibration (Dressel, 2019), without taking into account whether these are random or systematic or even known variations from one filter to the other. It must be noted that systematic errors, particularly those regarding absolute calibration, are usually substantially higher than random errors. The former ones can reach up to 10-20%, while the latter ones can be assumed to be always around 1%. This will be of interest later on, when the fitting algorithm is described in section 3.3.

The filters used in this work are F225W, F336W, F410M, F502N, F547M, F689M, FQ727N, FQ750N, FQ889N, FQ937N (where the 3-digit numbers in the filter name refer to each filter's effective wavelength) and the pixel scale of the images (300 km/pixel without binning, proportionally increased in quad filters for optimizing signal to noise and exposure time) are described in Dressel (2019). The ultraviolet filters are, generally speaking, sensitive to Rayleigh scattering by the atmospheric gas and to the properties of the sub-micron sized particles at the upper atmosphere. On the other hand, narrow filters covering methane absorption (FQ727N – intermediate, FQ889N – deep), when used together with near continua (FQ750N, FQ937N) are able to provide an altimetry of the cloud tops. A first estimation of the relative altitudes of the triple vortex, with the anticyclones bright at methane bands and dark at short wavelengths, and the opposite for the cyclone, provides a crude picture of the two anticyclones located higher than the cyclone. A similar behavior is observed in images taken with the Imaging Science Subsystem (ISS) onboard the Cassini spacecraft (del Rio-Gaztelurrutia et al., 2010; Vasavada et al., 2006) where cyclones appear bright in the BL1 filter and dark in the MT2 and MT3 filters. However, in order to retrieve a more detailed description of the vertical structure, we need to perform a detailed radiative transfer analysis of the data.

| Date | B | B' | α | Filters |
|---|---|---|---|---|
| **2015/06/29** | 28.74 | 29.62 | 3.65 | F225W, F336W, F410M, F502N, F547M, F689M, FQ727N, FQ750N, FQ889N, FQ937N |
| **2015/06/30** | 28.63 | 29.62 | 3.72 | F225W, F336W, F410M, F502N, F547M, F689M, FQ727N, FQ750N, FQ889N, FQ937N |
| **2015/07/01** | 28.73 | 29.63 | 3.79 | F225W, F336W, F410M, F502N, F547M, F689M, FQ727N, FQ750N, FQ889N, FQ937N |

Table 1: Observations: *B* (sub-earth planetocentric latitude); *B'* (sub-solar planetocentric latitude; *α* (phase angle).

*2.2. Data selection.*

From all the available data, we have selected a range of latitudes, from 55ºN to 69ºN where we can observe the atmospheric feature of interest and the structure of bands and zones in the surrounding background atmosphere.

Since we have images taken on different days, there is a longitude drift in the positions of individual features (such as the vortices) following the zonal wind profile (García-Melendo et al., 2011). We have considered such a drift, and studied the longitude box surrounding the triple vortex that is visible at the three visits, at least in one image for each of them. When more than one observation was available, the values of reflectivity and geometry were averaged, since the differences in observing conditions were small for each visit. This allowed us to obtain three spectra for every point of the region of interest, each one at a different viewing and illumination conditions (Figure 3).

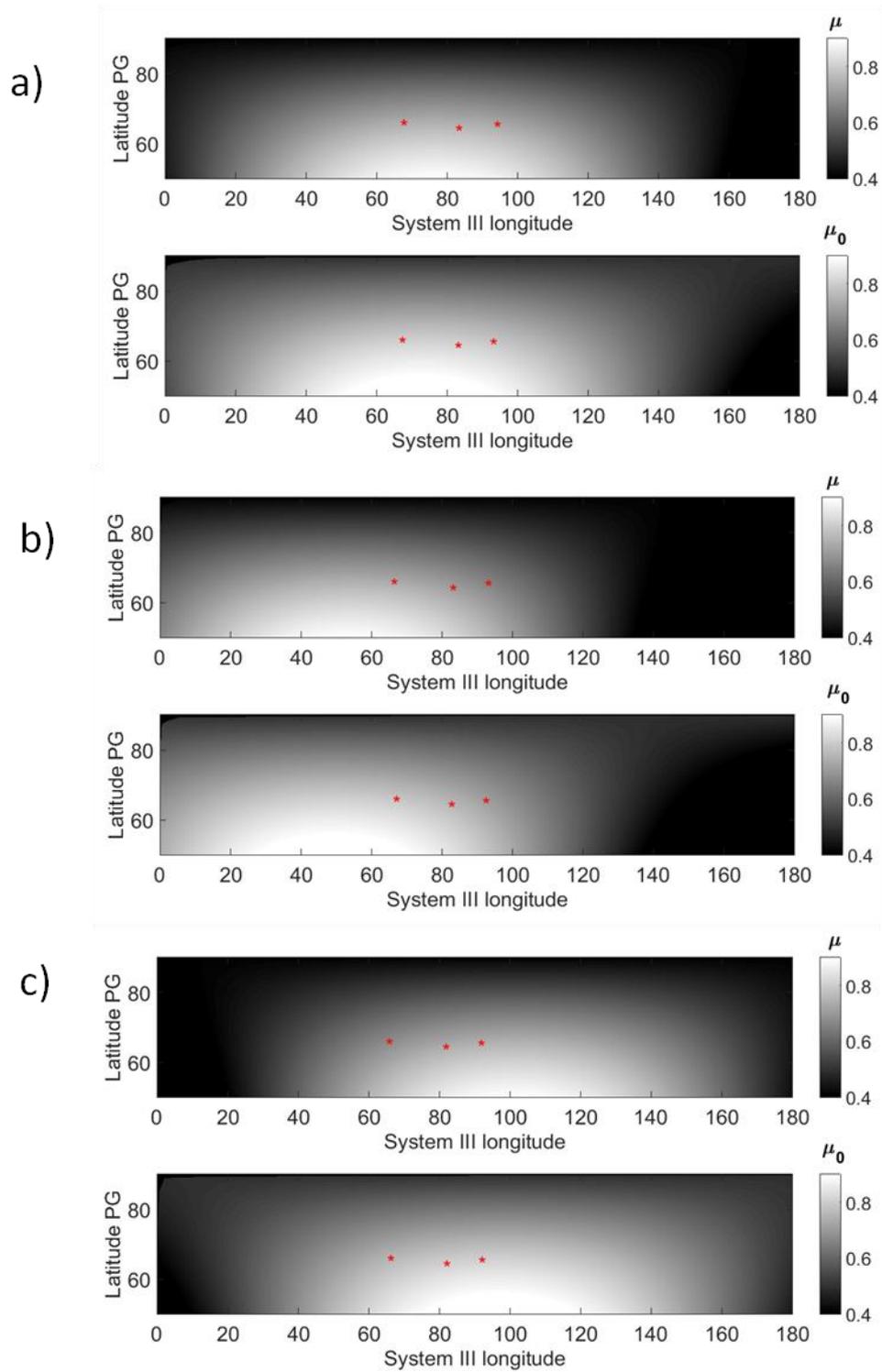

Figure 3: Maps for the values of the cosine of the emission ($\mu$) and emission ($\mu_0$) angles for the different dates of the images used in this work. a) June-29, b) June-30 c) July-1. The location of the triple vortex is indicated on each map with red stars.

In Figure 4 we show cylindrical projections of the region of interest in every filter. These images have been corrected for limb-darkening only for display purposes, as the limb-darkening information will be used in the following sections to constrain atmospheric properties. These images are averages from one or more original images, depending on the latitude and longitude coverage of the HST observations for each case.

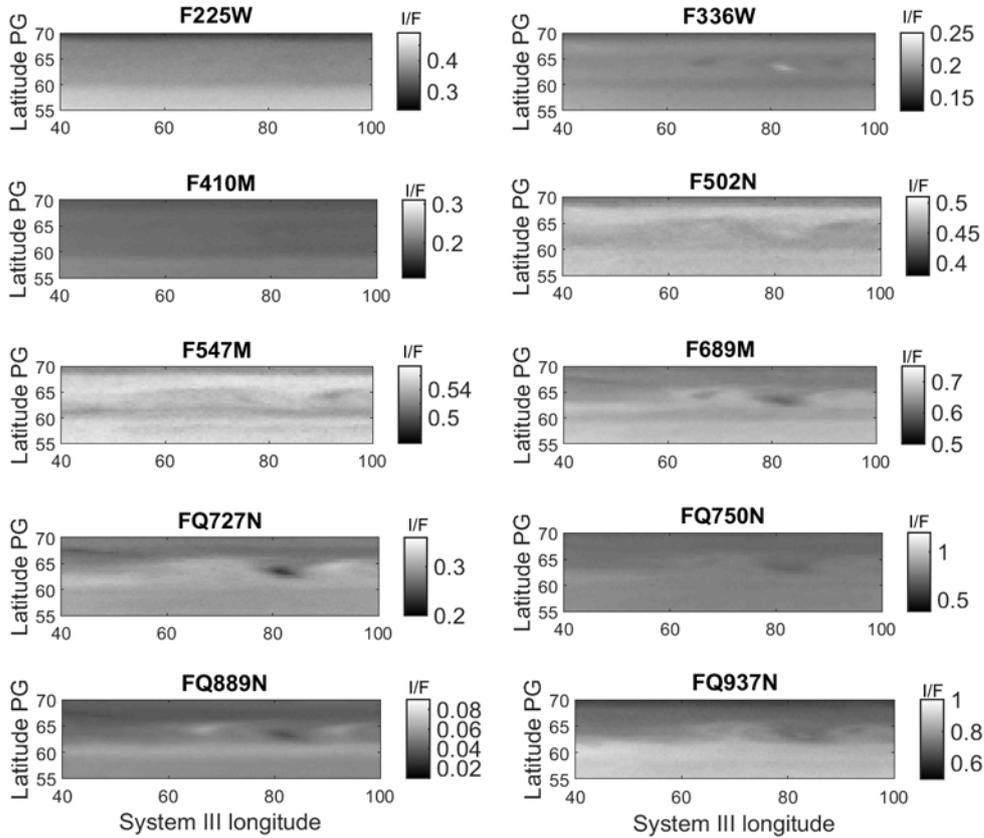

Figure 4: Cylindrical maps at the available wavelengths showing the region under study with A-C-A system included corrected for limb-darkening. The position of the triple vortex is centered at approximately 85ºE and is most apparent in filters F336W, FQ889N and FQ937N.

## 3. Methodology

*3.1 Radiative Transfer code*

Our goal is to reproduce the observed dependence of absolute reflectivity with geometry (three combinations of incidence and emission angles) for all wavelengths at the same time, so we can deduce the values of different parameters that give us information about the atmosphere such as the optical thicknesses of aerosol layers, the mean size of the particles, the height at which they are found the different layers and so on. To do this we used the radiative transfer code and retrieval suite NEMESIS, developed by Irwin et al.

(2008). This code uses the optimal estimator scheme to find the most likely model that best accounts for the observations.

The version of the code used here is based in a doubling-adding scheme that assumes a plane–parallel atmosphere to compute the emergent intensity of reflected sunlight due to scattering and absorption from atmospheric aerosols and gases. In our model we also take into account the Rayleigh scattering due to the mixture of $H_2$ and He, as well as the absorption due to $CH_4$. The general assumptions (temperature-pressure profile and gaseous abundances) used in this work are the same as in Sanz-Requena et al. (2018).

*3.2 Vertical cloud structure model*

Previous works (Pérez-Hoyos et el. 2005, Sanz-Requena et al. 2018) have found that a vertical structure consisting of three distinct layers of particles is good enough to reproduce the spectral and geometric variations of the absolute reflectivity at visible wavelengths. The overall vertical distribution of particles assumed in the present work is similar to that of Sanz-Requena et al., 2018, as shown in Figure 5. In Table 2, we summarize the list of free and fixed parameters, which have been chosen according to previous works (Pérez-Hoyos et el. 2016, Sanz-Requena et al. 2018). The same is true for the description of the gaseous scattering (by a mixture of $H_2$ and He, with a volume mixing ratio of 0.124 relative to $H_2$, (de Pater and Lissauer, 2001) and absorption. We only considered absorption by $CH_4$, using pre-computed k-tables based on the absorption coefficients given by Karkoschka and Tomasko (2010).

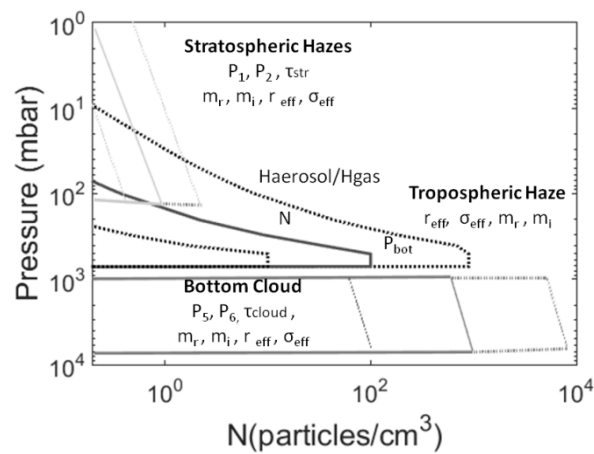

Figure 5: A priori assumed particle density profiles in our three assumed cloud/haze layers (solid lines) and their corresponding uncertainties (dotted lines). Parameters of the model for each layer are also indicated (see the text for a full explanation). Adapted from Sanz-Requena et al., 2018.

The uppermost aerosol layer corresponds to the stratospheric haze that is located between $P_1$ = 1 mbar and $P_2$ = 100 mbar (Pérez-Hoyos et al., 2005). In this layer, we assume a constant refractive index with real part and imaginary parts for all wavelengths, which we set to the average of that for ammonia ice ($m_r$ = 1.43 and $m_i$ = $10^{-3}$; Pérez-Hoyos et al., 2016). We also set the effective radius and the effective variance to be 0.1 μm and 0.1, respectively, and assume that the particle size distribution follows a log-normal distribution (Hansen and Travis, 1974). The only free parameter in this layer is the optical thickness, for which we have set a starting point $\tau_{str}$= 0.01±0.01 (Sanz-Requena et al., 2018), at 900 nm, which will be used as the reference wavelength in the following analysis, except where stated otherwise.

The second layer, corresponding to the tropospheric haze, is characterized by a variable optical thickness ($\tau_{trop}$ = 10 ± 2) (Karkoschka and Tomasko, 2005), as well as a parameterization of its vertical distribution. This is defined by the pressure corresponding to the lower base (600 ± 100 mbar; Fletcher et al., 2007, Roman et al., 2013). The particle-to-gas scale height ratio of this aerosol layer is taken initially as $H_{aerosol}/H_g$ = 0.7 ±0.1 (Pérez-Hoyos et al., 2016) and the value of the initial maximum concentration of particles is N = 20 ± 10 particles/cm$^3$ (Sanz-Requena et al., 2018). Here $H_g$ is the atmospheric (gas) scale height ∼ 38 km.

We assume that the particles are spherical and we use Mie theory to compute the phase function. Since all the observations are made at a similarly low phase angle value (∼ 3.6º), this assumption is not critical. We have taken the initial values of $r_{eff}$ = 1.5 ± 0.5 and $\sigma_{eff}$ = 0.1 ± 0.1. (Ortiz et al., 1996). Our model calculates the real refractive index from Kramers-Kronig's relation (Lucarini et al., 2005) from an initial value $m_r$ = 1.43. The imaginary refractive index is set as a free parameter taking as initial value $m_i$ = $10^{-3}$ ± $10^{-3}$ for all wavelengths (Roman et al., 2013).

The lower layer is fixed between pressures $P_5 = 1.0$ bar and $P_6 = 1.4$ bar and corresponds to the cloud putatively formed by ammonia ice (Roman et al., 2013). It must be noted that such ammonia ice has been very rarely spectroscopically identified, with a few exceptions (Baines et al., 2009; Sromovsky et al., 2015). The refractive indices are fixed ($m_r = 1.43$ and $m_i = 10^{-3}$) as in the stratospheric haze, and the optical thickness is the only free parameter (with a priori values of $\tau_{cloud} = 10 \pm 2$ from Pérez-Hoyos et al., 2016). As in stratospheric haze, the particle size distribution is log-normal with an effective radius of 10 µm and an effective variance of 0.1 µm (West et al., 2009).

| Layer | Parameter | Type | Value |
|---|---|---|---|
| **StratosphericHaze** | $P_1$ | Fixed | 1 mbar |
|  | $P_2$ | Fixed | 100 mbar |
|  | $\tau_{str}$ | Free | 0.01±0.01 |
|  | $m_r$ | Fixed | Amonnia ice |
|  | $m_i$ | Fixed | Amonnia ice |
|  | $r_{eff}$ | Fixed | 0.1 µm |
|  | $\sigma_{eff}$ | Fixed | 0.1 µm |
| **Tropospheric Haze** | $P_{bot}$ | Free | 600±100 mbar |
|  | N | Free | 20 ±10 particles/cm3 |
|  | H | Free | 25±5 km |
|  | $\tau_{trop}$ | Computed | 10±5 |
|  | $r_{eff}$ | Free | 1.5±0.5 µm |
|  | $\sigma_{eff}$ | Free | 0.1 ±0.1µm |
|  | $m_r$ | Fixed | 1.43 |
|  | $m_i(225\ µm)$ | Free | $10^{-3}\pm10^{-3}$ |
|  | $m_i(336\ µm)$ | Free | $10^{-3}\pm10^{-3}$ |
|  | $m_i(410\ µm)$ | Free | $10^{-3}\pm10^{-3}$ |
|  | $m_i(502\ µm)$ | Free | $10^{-3}\pm10^{-3}$ |
|  | $m_i(547\ µm)$ | Free | $10^{-3}\pm10^{-3}$ |
|  | $m_i(689\ µm)$ | Free | $10^{-3}\pm10^{-3}$ |
|  | $m_i(727\ µm)$ | Free | $10^{-3}\pm10^{-3}$ |
|  | $m_i(750\ µm)$ | Free | $10^{-3}\pm10^{-3}$ |
|  | $m_i(889\ µm)$ | Free | $10^{-3}\pm10^{-3}$ |
|  | $m_i(937\ µm)$ | Free | $10^{-3}\pm10^{-3}$ |
| **Bottom Cloud** | $P_5$ | Fixed | 1.0 bar |
|  | $P_6$ | Fixed | 1.4 bar |
|  | $\tau_{cloud}$ | Free | 10±5 |
|  | $m_r$ | Fixed | 1.43 |
|  | $m_i$ | Fixed | $10^{-3}$ |
|  | $r_{eff}$ | Fixed | 10µm |
|  | $\sigma_{eff}$ | Fixed | 0.1 µm |

Table 2: Model atmosphere parameters. Please note that $\tau_{trop}$ is in fact computed from the other parameters describing the vertical distribution.

*3.3. Fitting strategy*

To estimate the goodness of fit between the observed and modelled reflectivities, we evaluated the error function $\chi^2/n$ at every point of the free-parameter space. The error function is defined at each filter observation as:

$$\frac{\chi_\lambda^2}{n} = \frac{1}{n}\sum_{i=1}^{n}\frac{1}{\sigma_i^2}[(I/F)_{obs} - (I/F)_{mod}]^2$$

where $n$ is the number of points to be fitted by the model (e.g., the number of points scanning over longitude at a given latitude); $\sigma_i$ is the error in the ith measurement; and $(I/F)_{obs}$ is the observed and $(I/F)_{mod}$ the modeled reflectivity at a given point. Then we calculate a reduced $\chi^2/n$, an average of all the filters and positions over the disk being modeled. When $\chi^2/n$ is larger than one, the profiles deviate systematically from the data. When $\chi^2/n$ is smaller than 1, we can accept the model as it is, on average, inside the data error bars. Our goal is to find models close enough to the data for all geometries and wavelengths ($\chi^2/n < 1$) at every point of the region of interest. This approach does not take into account any differences between systematic and random errors and only minimizes the overall model deviation from the data. As low pixel-to-pixel noise allows determining limb-darkening more precisely than absolute reflectivity values (mostly affected by systematic errors), we will discuss in section 4.1 how well best-fitting models are able to match the observed limb-darkening.

We first fitted center-to-limb brightness profiles for the range of selected latitudes 55º to 69ºN with intervals of 0.5º, excluding the regions where the triple vortex is present. In this way, we obtain a reference model that we use later, including the regions where the triple vortex is located. We initially fitted filters FQ727N, FQ750N, FQ889N and FQ937N with a constant value of $m_i = 10^{-3}$. Once the result was acceptable we added the shortest wavelengths (filters F225W, F336W, F410M, F502N, F547M and F689M) leaving also the refractive indices as free parameters. This procedure avoids overfitting with the imaginary refractive index (whose values are forced to be nearly flat for longer wavelengths), provides good results and it is very similar to the strategy in Sanz-Requena et al. (2018). The result of this analysis is presented in Fig. 6, where we show the ten filters covering a wide geometrical range.

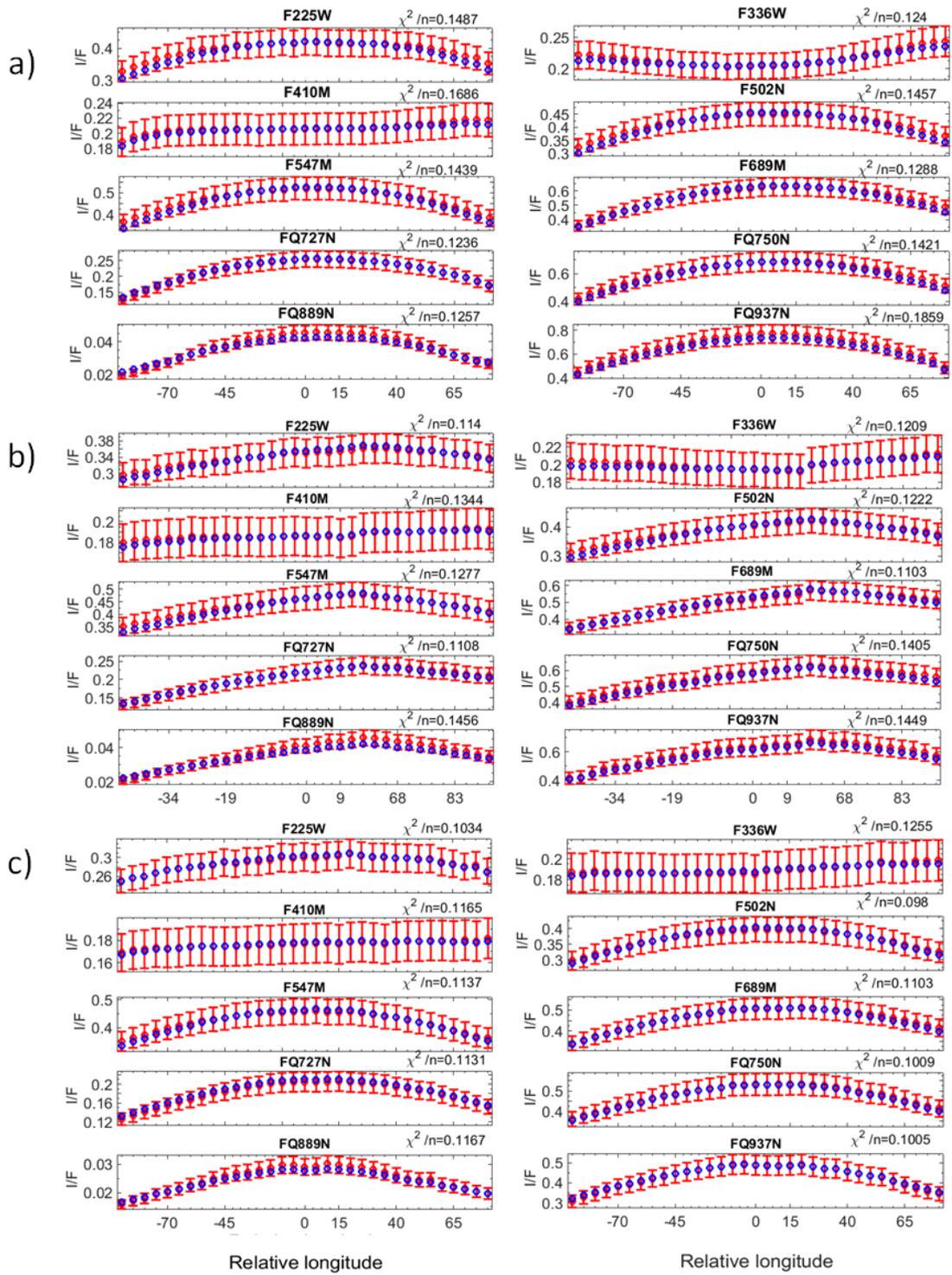

Figure 6: Best-fitting models for center-to-limb variations of reflectivity. Blue circles represent the modeled values and the red circles and lines correspond to the observed reflectivity and its corresponding error bar. (a) latitude 56ºN , (b) latitude 61ºN , and (c) latitude 69ºN . In this figure, longitudes are measured in degrees from an arbitrary reference longitude.

Once we have a good model for the overall limb-darkening behavior of the reference atmosphere, we want to fit individually all the points in the region of interest, including particular features such as the vortices. For doing so we run new retrievals using as input the reference atmosphere model at a given latitude. In figure 7 we show best-fitting model results for each of the three vortices at some locations of interest.

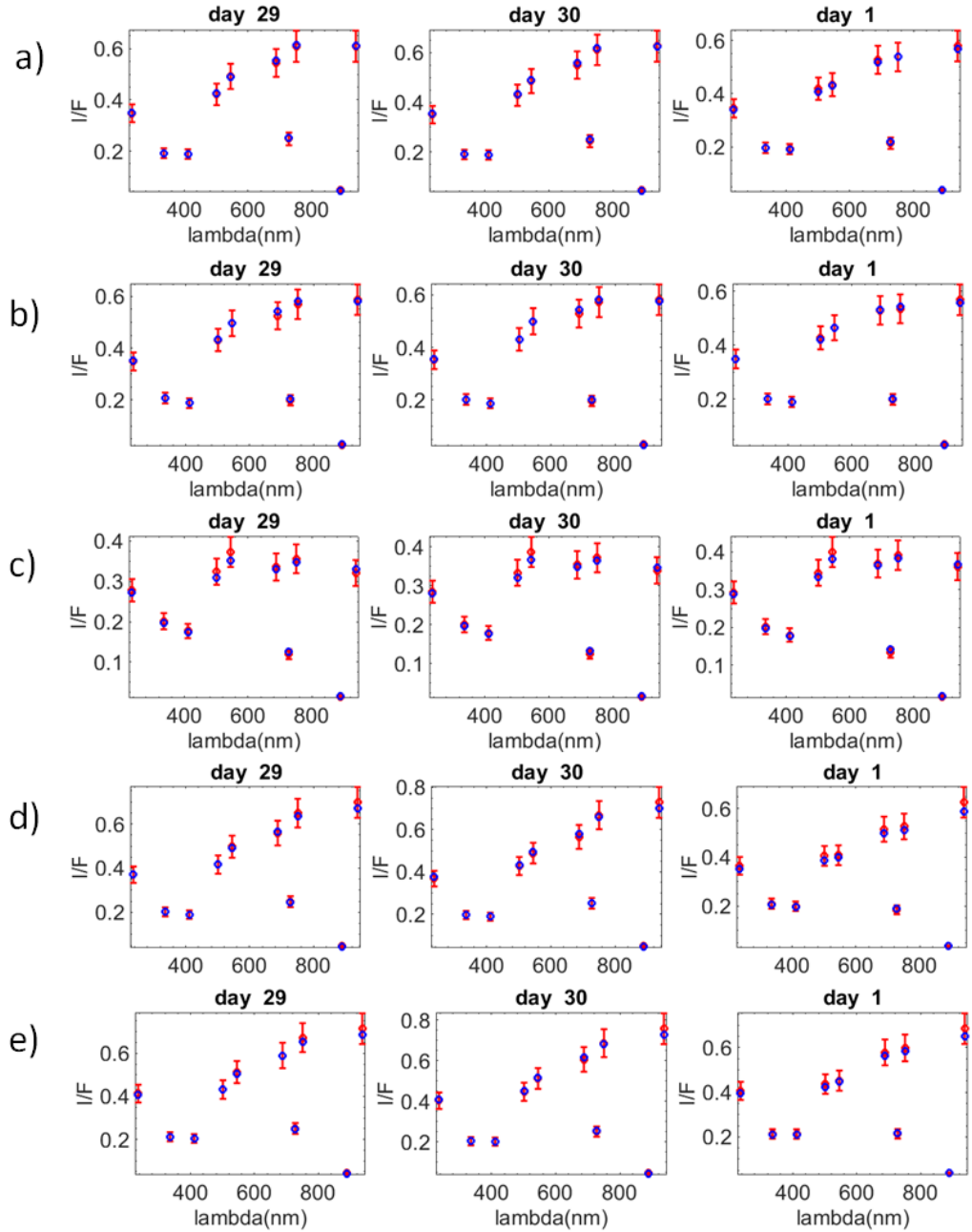

Figure 7: Examples of best-fitting results. Blue circles indicate the modeled values and the red circles and error bars correspond to the observed reflectivities and their corresponding error bars. Panel a) show longitude 60º- latitude 68º, panel b) show longitude 66º- latitude 64º (anticyclone), panel c) show longitude 81º - latitude 63º (cyclone), panel d) show longitude 50º - latitude 58º and panel e) show longitude 60º - latitude 64º.

## 4. Modelling Results

*4.1. General results and limb-darkening*

We have assumed error bars for the data that do not separate random from systematic errors. In order to rule out and quantify a possible systematic deviation of the limb-darkening in the different regions and filters and in order to improve our model we fitted both observations and best- fitting models to a Minnaert law.

$$\frac{I}{F} = \left(\frac{I}{F}\right)_0 \mu_0^k \mu^{k-1}$$

Where $(I/F)_0$ is the reference reflectivity value at perfect nadir geometry ($\mu=1$ and $\mu_0=1$) and $k$ is a wavelength-dependent limb-darkening coefficient (Sanz-Requena et al., 2018). This analysis corresponds to the background models where center-to-limb data is available, and not to the specific cyclonic features. While previous figure 7 explicitly shows the deviation between models and data, we show in Figure 8 the results corresponding to the variation of the values of $k$ for the different latitudes and for all wavelengths. As previously stated, the limb-darkening can be more precisely determined from the original data, as it is mostly affected by pixel-to-pixel and random errors, which are substantially lower than systematic calibration uncertainties. Although it would be desirable to be able to fit independently the limb-darkening with stronger constraints, our current version of the retrieval code does not have this capability. We want to show here that, at least, our best-fitting models are close enough to the measured limb-darkening values.

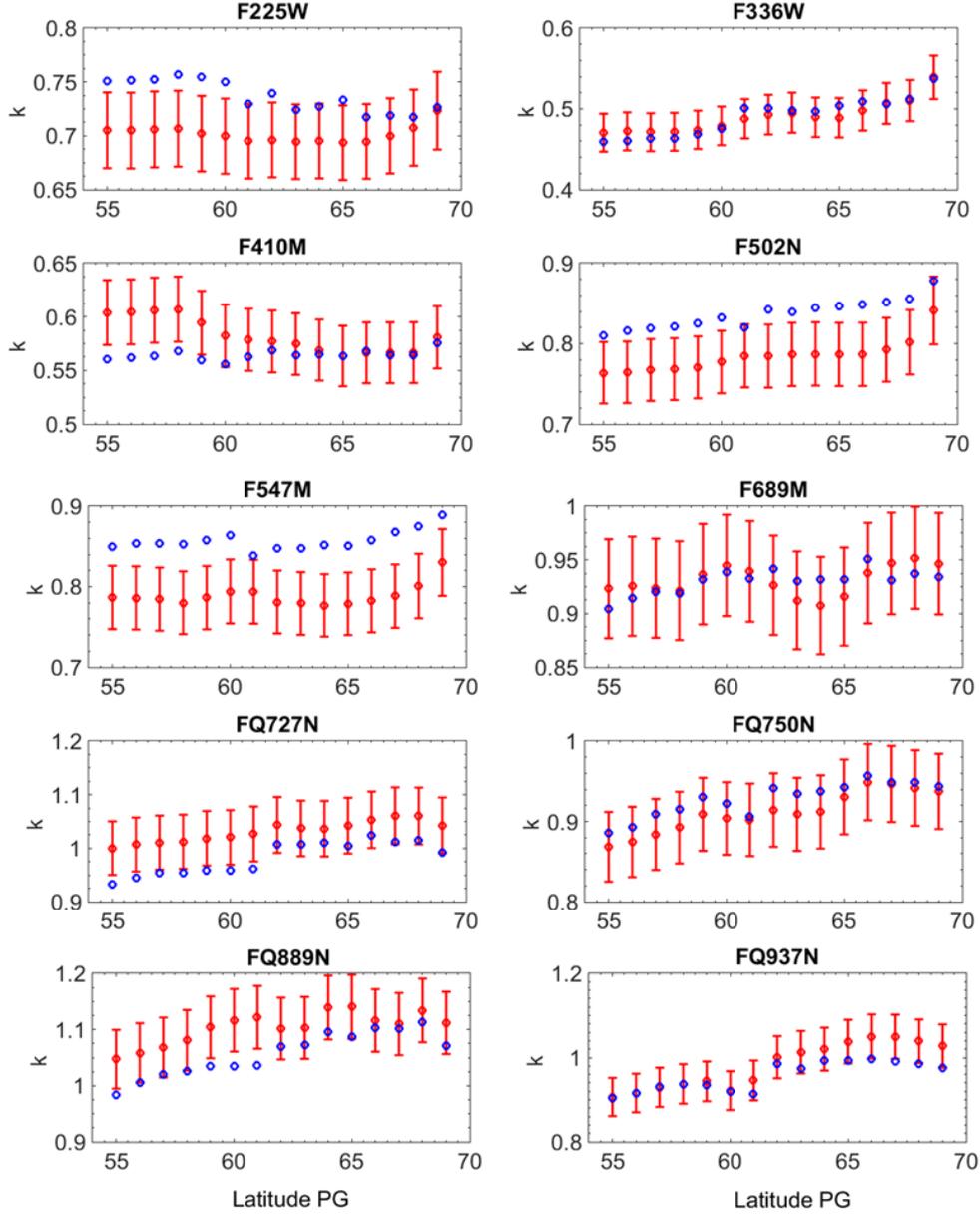

Figure 8: Values of the limb darkening coefficient *k* for all latitudes (55ºN to 69ºN and filters (F225W, F336W, F410M, F502N, F547M, F689M, FQ727N, FQ750N, FQ889N, FQ937N). Blue circles correspond to the best fits and the red circles and lines correspond to the HST data and their error bar. Error bars for the observed limb-darkening is taken as a 5% of the value of the HST data. This includes not only pixel-to-pixel random noise but also navigation uncertainties.

There is no systematic deviation in the limb-darkening of the models from that of the data, as the same filter or region can give differences as low as 0.15 % or as high as 7%, with average differences of 3%. Taking into account not only the errors in relative photometry but also the navigation uncertainties (including the longitudinal drift due to zonal winds) the level of discrepancy between observations and models is acceptable and indicates that our model atmosphere reproduces well the observed limb-darkening.

*4.2. Best-fitting results*

In figure 9 we show the goodness of the fit ($\chi^2/n$) for all the points of the region of study. We find that at all points $\chi^2/n <1$. However, there is a longitude trend in the goodness of fits, which implies that the overall limb-darkening behaviour is not perfectly constrained when we study each point of the grid separately. This implies that the state derived from the limb-darkening analysis is not sufficiently well constrained and the optimal results retrieved from a point-by-point basis do not fully reproduce the observed limb-darkening, as we will discuss later.

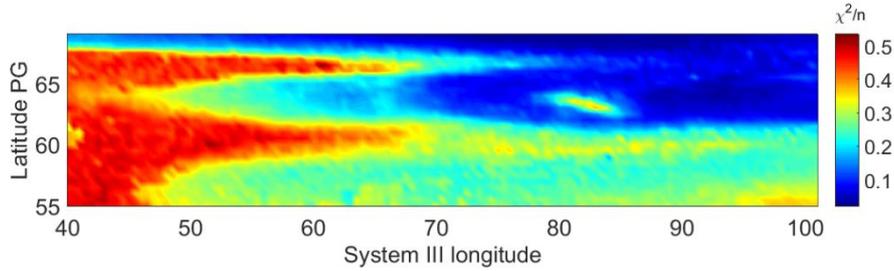

Figura 9: Reduced value of $\chi^2/n$ for all the points of the grid. All the fittings are satisfactory although limb-darkening seems not to be perfect in our models, as there is a longitude trend in the goodness of fit.

Figures 10 to 14 show the best-fitting parameters as a function of latitude and longitude. The optical thickness (at 0.9 μm) of the stratospheric haze (Figure 10a) shows no zonal variation and depends mostly on latitude. We find an increase with latitude that varies from $\tau_{str} = 0.01 \pm 0.01$ (~55º N) to $\tau_{str} = 0.025 \pm 0.01$ (~69ºN). This implies an increase of a factor 3 in the stratospheric particle density from the lower to the upper latitudes, similar to that presented in Sanz-Requena et al. (2018). It is important to notice that the models do not require differences in the thickness of the upper haze between the anticyclone-cyclone system and the surrounding regions ($\tau_{str}$ ~ $0.015 \pm 0.01$).

In Figure 10b we show the optical thickness down to the bottom cloud at the ammonia condensation levels. In this case we find that the values are quite homogeneous, with hardly any spatial variation ($\tau_{cloud}$ ~ $9.2 \pm 2$ to ~55ºN and $\tau_{cloud}$ ~ $8.4 \pm 2$ to ~ 69ºN).

The pressure level for the base of the tropospheric haze (Figure 10c) varies from 700 ± 100 mbar for ~ 55ºN to 300 ± 100 mbar for ~ 69ºN. The values are very homogeneous in longitude. However, there are small differences in the region of the triple vortex. The value in the anticyclones is approximately 550 ± 100 mbar while in the cyclone region the average value is 500 ± 100 mbar. This implies that in terms of altitude above the 1 bar level, the tropospheric haze is located at 40 ± 5 km and 50 ± 5km for the anticyclones and the cyclone respectively. On the other hand, as we approach higher latitudes the height of the base of the tropospheric haze increases from 20 ± 5 km at 55ºN to 60 ± 5km at 70ºN .

The maximum particle concentration (Figure 10d) follows a latitudinal behavior similar to that of the base pressure of the tropospheric haze. Its value ranges from 100 ± 10 particles /cm$^3$ for 55ºN to 50 ± 10 particles/cm$^3$ for 70ºN. Again, the behavior is quite homogeneous in longitude, with small differences at the anticyclones (80 ± 10 particles/cm$^3$) and the cyclone (55 ± 10 particles/cm$^3$), while in the cyclone we can see a decrease in the maximum average concentration.

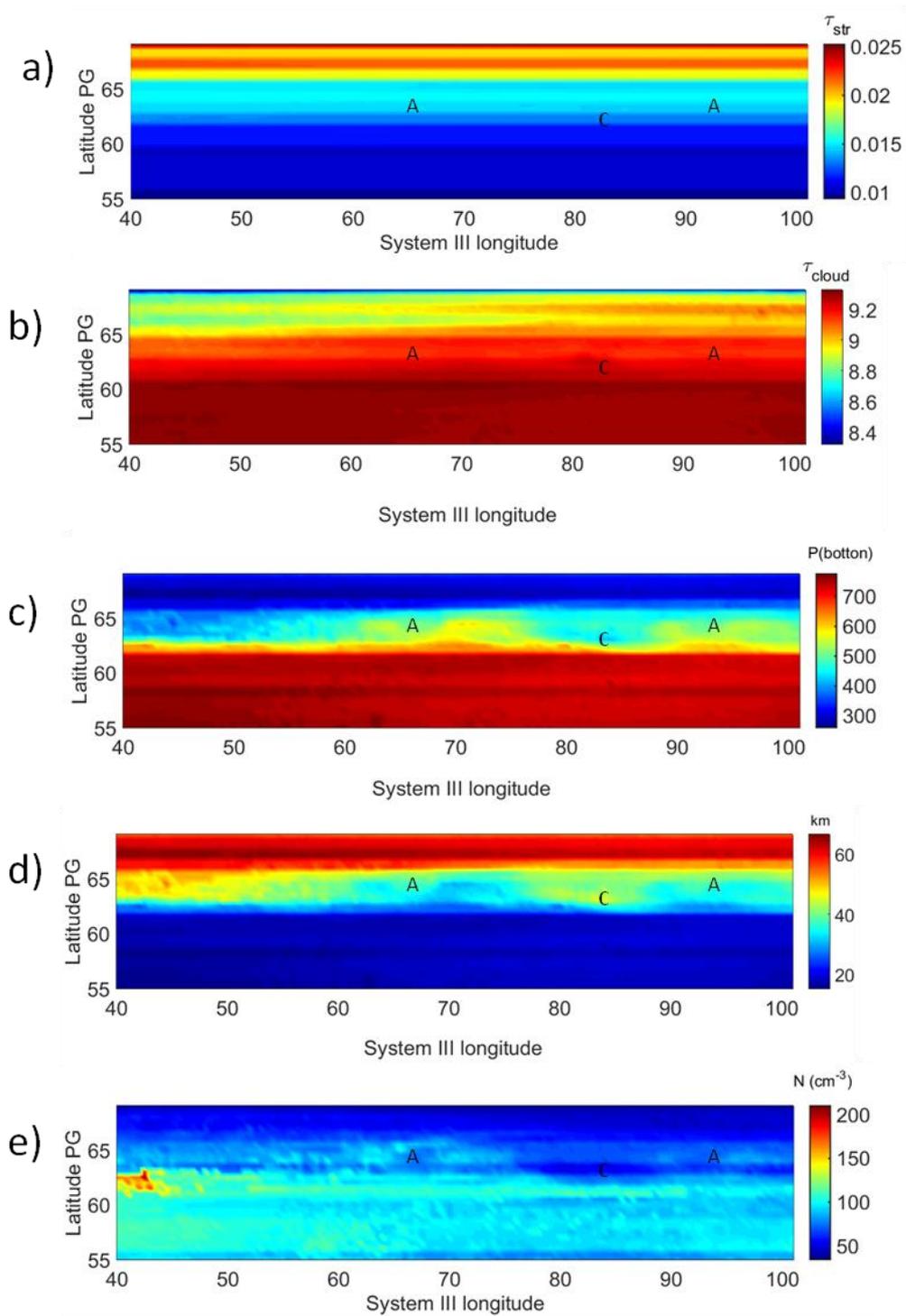

Figure 10: a) Optical thickness of the stratospheric haze. b) Optical thickness down to the bottom cloud. c) Pressure (mbar) of the base of the tropospheric haze. d) Height (km) of the base of the tropospheric haze. e) Maximum particle concentration (particles/cm$^3$). The location of the triple vortex is indicated on each map. (Anticyclone (A), Cyclone (C), Anticyclone (A))

In Figure 11a we show the variation of particle density with height for 6 different regions. We observe that for low latitudes (58ºN ) maximum concentrations (~ 110 ± 10

particles/cm$^3$) are located at pressures of ~ 900 ± 100 mbar. The maximum concentrations in the two anticyclonic regions (64ºN and 65 ºN) have similar values (~85 ± 10 particles / cm$^3$) and both are at the same pressure level (~550 ± 100 mbar). In the cyclonic region (63ºN) we observe a smaller peak concentration (~ 55 ± 10 particles / cm$^3$) at a lower pressure (~ 500 ± 100 mbar). Outside the triple vortex, we find that at the same latitudes the maximum concentration (~ 70 ± 10 particles /cm$^3$ and ~600 ± 100 mbar is smaller than that of the anticyclones and bigger than that of the cyclone. The value of the maximum concentration at higher latitudes (68ºN) is ~ 60 ± 10 particles /cm3 and is located at a pressure similar to that of the cyclone.

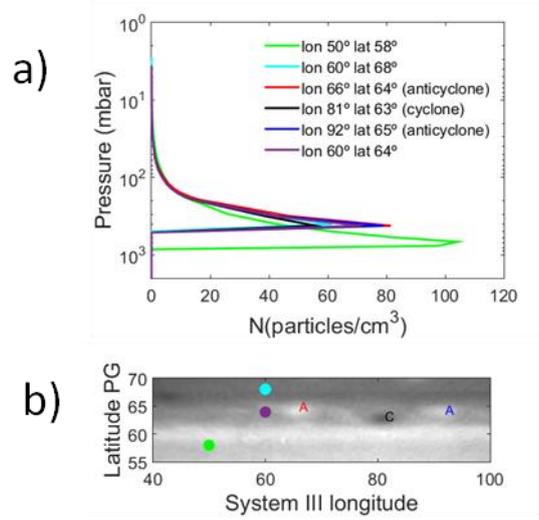

Figure 11: a) Vertical distribution of the tropospheric particles for the vortices and a reference region. b) Location of vortices and reference regions

Figure 12a shows the fractional scale height of the tropospheric haze, $H$aerosol/$H$gas. We observe a decrease of this parameter with latitude, 0.65 ± 0.1 ($H$aerosol ~ 23± 2 km) for ~ 55ºN and 0.45 ± 0.1 ($H$aerosol ~16± 2 km) for ~69ºN . In the latitudes where the triple vortex is found, we do not observe substantial differences between the values corresponding to the anticyclones and the cyclone (0.55 ± 0.1) ($H$aerosol ~ 18± 2 km).

The optical thickness of the tropospheric haze is also quite homogeneous in longitude except at the triple vortex (Figure 12b). In latitude we observe an increase of optical thickness from $\tau$trop ~ 28± 2 at 55ºN to a maximum $\tau$trop ~ 35± 2 at 61º N , and then the magnitude decreases down to $\tau$trop ~ 10± 2 at 69ºN . This is consistent with the belt and zone structure of the region. The values of the optical thickness in the anticyclones is $\tau$trop ~ 25± 2, similar to the average at their latitude, while the optical thickness of the cyclone it is ~ 10± 2. As we will see, this parameter accounts for most of the spectral and geometrical variation in this data set.

We have calculated the vertical thickness of the tropospheric haze (Figure 12c) from the height corresponding to the optical thickness equal to 1 down to the base level, as a proxy to the vertical extension of the haze. We observe that the tropospheric haze thickness decreases polewards from 80 ±5 km at ~55º N to 40 ±5 km at ~69ºN. Again, the behavior in longitude is quite homogeneous. However, we do find differences between the anticyclonic region and the cyclonic region with thicknesses of 60±5 km and 50±5 km respectively.

Regarding the particles size, we do not find significant differences for the range of latitudes that we are considering, being ~ 0.15 ± 0.1 μm. (Figure 12d), with a subtle belt/zone structure that is inside the parameter error bars and thus not statistically significant.

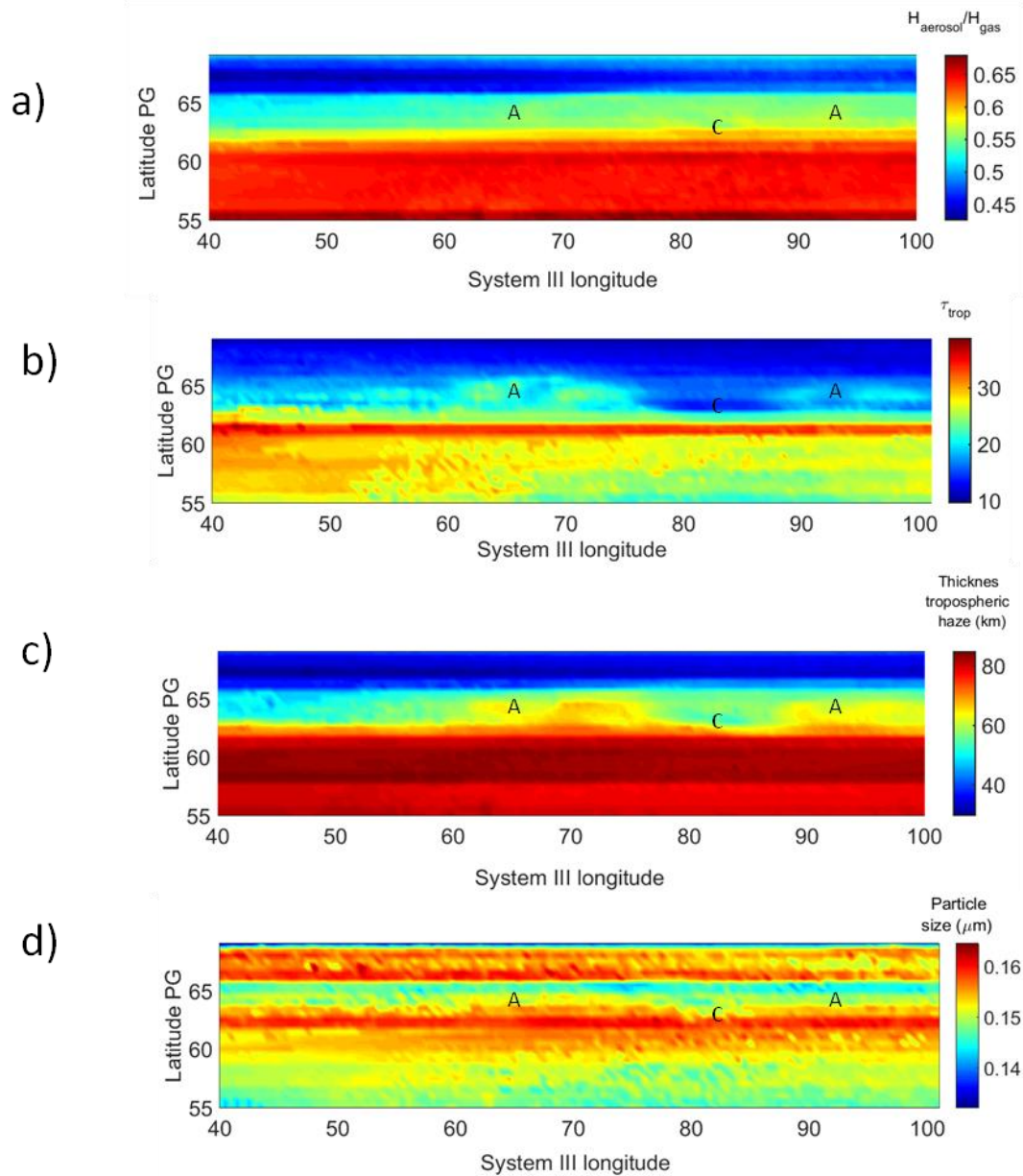

Figure 12: a) Scale height of the tropospheric haze, ($H_{aerosol}/H_{gas}$). b) Optical thickness of the tropospheric haze. c) Vertical thickness of the tropospheric haze (km). d) Particle size (μm). The location of the triple vortex is indicated on each map. (Anticyclone (A), Cyclone (C), Anticyclone (A))

Figure 13 shows the imaginary refractive indices of the tropospheric haze for six different wavelengths. We observe that the values are quite homogeneous in longitude. We have omited the values for wavelengths 725 nm, 750 nm, 889 nm and 937 nm since they are practically constant. We do not appreciate significant differences at the locations of the anticyclones and the cyclone at any wavelength. The relative errors of all parameters retrievals are displayed in Figure 14.

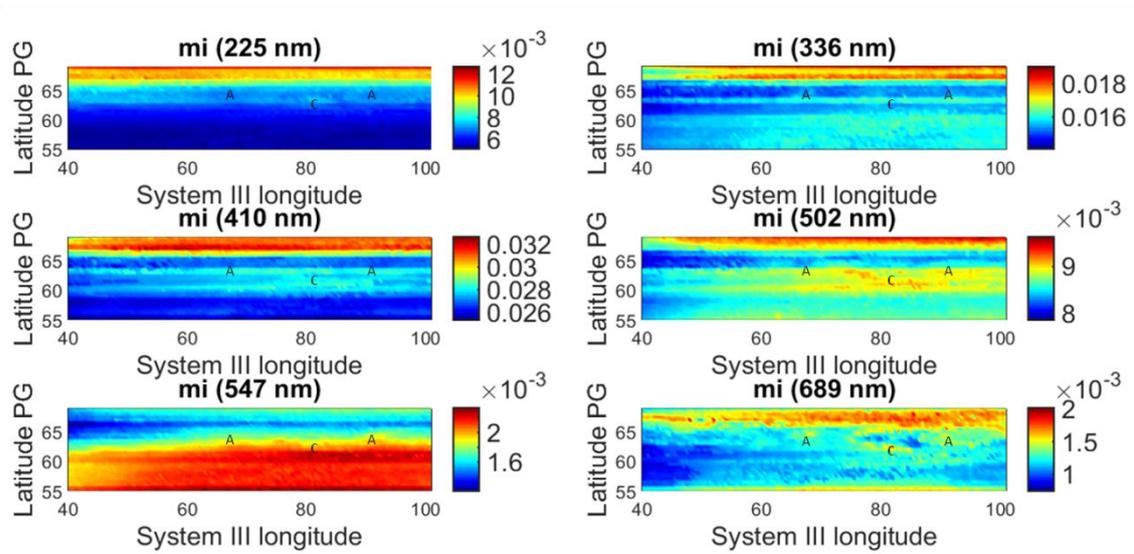

Figure 13: Imaginary refractive indexes for the tropospheric haze retrieved for the first six filters wavelengths. The location of the triple vortex is indicated on each map. (Anticyclone (A), Cyclone (C), Anticyclone (A))

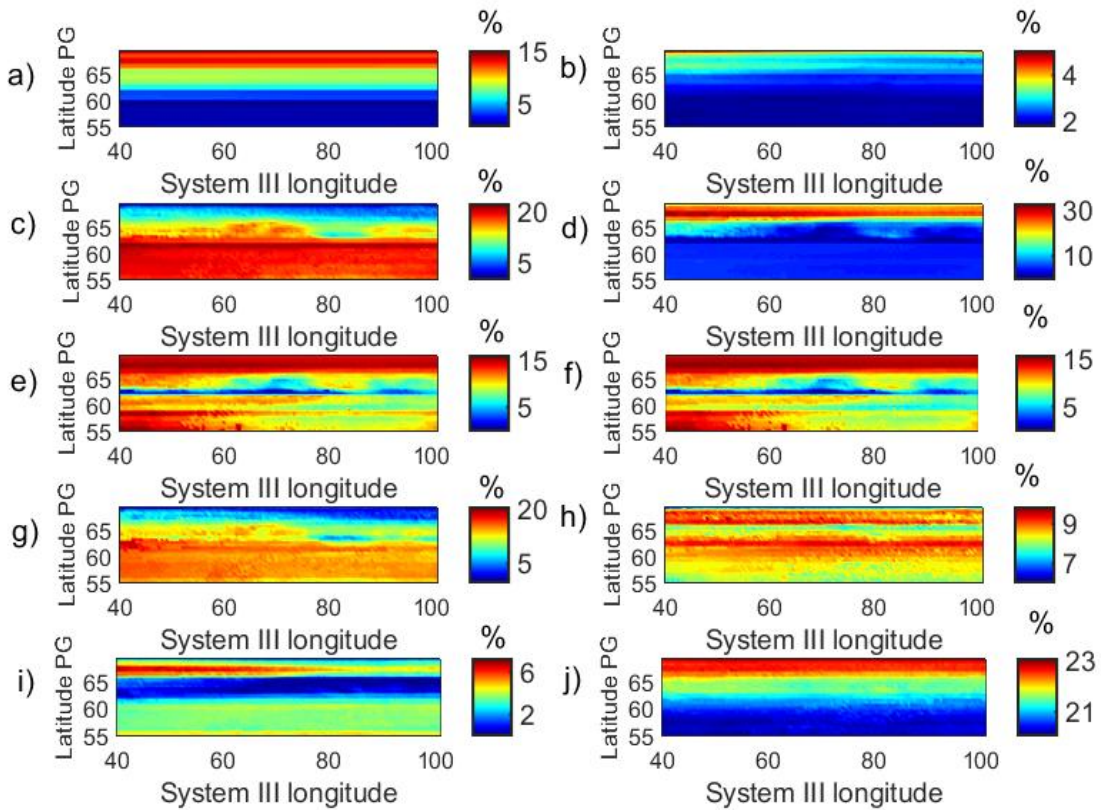

Figure 14: Relative errors. a) Optical thickness of the stratospheric haze. b) Optical thickness down to the bottom cloud. c) Optical thickness of the tropospheric haze. d) Pressure of the base of the tropospheric haze. e) Height of the base of the tropospheric haze. f) Vertical thickness of the tropospheric haze. g) Maximum particle concentration . h) Particle size, i) Scale height of the tropospheric haze, ($H_{aerosol}/H_{gas}$). j) Imaginary refractive index (410 nm)

In Figure 15, we show the variation of the imaginary refractive indexes $m_i$ with wavelength for six selected regions, including the locations of the cyclone and anticyclones. The behavior at all the six regions is similar. At visible and infrared wavelengths, $m_i$ decreases with wavelength, from ~ $28 \pm 0.1 \: 10^{-3}$ (410 nm) to $5 \pm 0.1 \: 10^{-4}$ (937 nm). In the shortest wavelengths, it increases from ~$7 \pm 0.1 \: 10^{-3}$ (225 nm) to ~$16 \pm 0.1 \: 10^{-3}$ (336 nm) and ~ $28 \pm 0.1 \: 10^{-3}$ (410 nm). In this range of wavelegths, $m_i$ is slightly higher at higher latitudes.

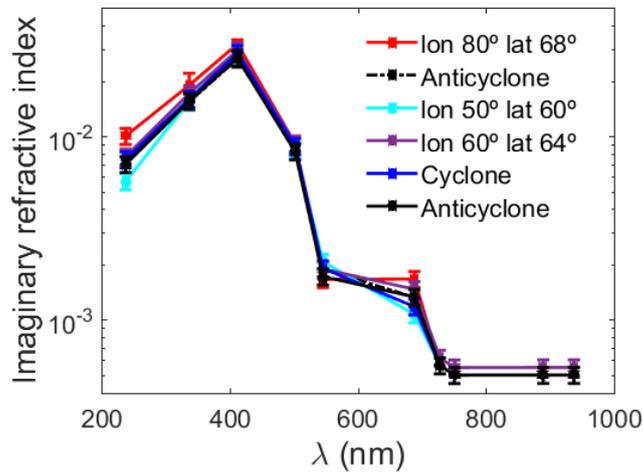

Figure 15: The imaginary refractive indexes respect to the wavelength for six selected regions.

Table 3 shows the best-fitting values of the parameters for different regions.

|  | anticyclone | cyclone | Region 1 | Region 2 |
|---|---|---|---|---|
| **Stratospheric Haze** | | | | |
| $\tau_{str}$(0.9 μm) | 0.015±0.01 | 0.015±0.01 | 0.01±0.01 | 0.025±0.01 |
| **TroposphericHaze** | | | | |
| z(km) | 40±5 | 50±5 | 20±5 | 60±5 |
| $P_{bot}$(mbar) | 550±50 | 500±50 | 750±50 | 300±50 |
| N(part/cm3) | 78±10 | 55±10 | 100±10 | 50±10 |
| H(km) | 18±2 | 18±2 | 23±2 | 17±2 |
| $\tau_{trop}$(0.9 μm) | 25±2 | 10±2 | 28±2 | 10±2 |
| Thicknes tropospheric haze(km) | 70±2 | 50±2 | 78±2 | 35±2 |
| $r_{eff}$(μm) | 0.14±0.1 | 0.15±0.1 | 0.15±0.1 | 0.15±0.1 |
| $\sigma_{eff}$(μm) | 0.05±0.01 | 0.05±0.01 | 0.05±0.01 | 0.05±0.01 |
| $m_i$(225 μm) | 7±0.1E-03 | 7±0.1E-03 | 6±0.1E-03 | 12±0.1E-03 |
| $m_i$(336 μm) | 15±0.1E-03 | 15±0.1E-03 | 16±0.1E-03 | 18±0.1E-03 |
| $m_i$(410 μm) | 26±0.1E-03 | 26±0.1E-03 | 26±0.1E-03 | 32±0.1E-03 |
| $m_i$(502 μm) | 7±0.1E-03 | 7±0.1E-03 | 8±0.1E-03 | 9±0.1E-03 |
| $m_i$(547 μm) | 1.7±0.1E-03 | 1.7±0.1E-03 | 3±0.1E-03 | 1.6±0.1E-03 |
| $m_i$(689 μm) | 1.3±0.1E-03 | 1.3±0.1E-03 | 1.3±0.1E-03 | 1.7±0.1E-03 |
| $m_i$(727 μm) | 5±0.1E-04 | 5±0.1E-04 | 5±0.1E-04 | 5±0.1E-04 |
| $m_i$(750 μm) | 5±0.1E-04 | 5±0.1E-04 | 5±0.1E-04 | 5±0.1E-04 |
| $m_i$(889 μm) | 5±0.1E-04 | 5±0.1E-04 | 5±0.1E-04 | 5±0.1E-04 |
| $m_i$(937 μm) | 5±0.1E-04 | 5±0.1E-04 | 5±0.1E-04 | 5±0.1E-04 |
| **Cloud** | | | | |
| $\tau_{cloud}$(0.9 μm) | 9±2 | 9.5±2 | 9.5±2 | 8.8±2 |

Table 3: Region 1 corresponds to longitude 80º - latitude 60ºN and region 2 corresponds to longitude 80º- latitude 68ºN.

According to our results, the behavior of the parameters is quite homogeneous in longitude both in the stratospheric haze and at cloud level. This same behavior is found in the tropospheric haze, except in the latitudes where the triple vortex appears.

*4.3. Sensitivity to the Model Parameters.*

It is possible to evaluate the information gain during the retrieval process by comparing the relative errors between the a priori assumption and the a posteriori best-fitting value. For doing so, we evaluate the improvement factor as defined by Irwin et al. (2015). A low improvement factor indicates that the a posteriori result of the free parameter is not giving us substantial information regarding the a priori uncertainty, while a high improvement factor indicates that we have significantly reduced the a priori uncertainty during the retrieval. In table 4 we show the results of the improvement factors for the different free parameters.

|  | improvement factor |
|---|---|
| **Stratospheric Haze** | |
| $\tau_{str}(0.9\ \mu m)$ | 2% |
| **TroposphericHaze** | |
| $P_{bot}$(mbar) | 92% |
| N(part/cm3) | 95% |
| Haerosol/Hgas | 15% |
| $r_{eff}(\mu m)$ | 12% |
| $\sigma_{eff}(\mu m)$ | 3% |
| $m_i$(lambda) | 10% |
| **Cloud** | |
| $\tau_{cloud}(0.9\ \mu m)$ | 5% |

Table 4: Improvement factor of free parameters.

According to these results, we can assess that the base altitude and peak concentration, and the scale height to a lesser extent are the most important parameters, and the retrieval more informative about their values.

To address the importance of these most informative free parameters in both the nadir-viewing reflectivity and the limb-darkening, we made a new Minnaert fit to models computed at nominal values as well as 1-σ above and below the nominal result. In figure 16 we show how the average value of the values of $I/F_0$ and of $k$ for the different wavelengths varies as a given parameter is changed. From these results we observe that the most affected filters are F336W, FQ750N, FQ889N and FQ937N.

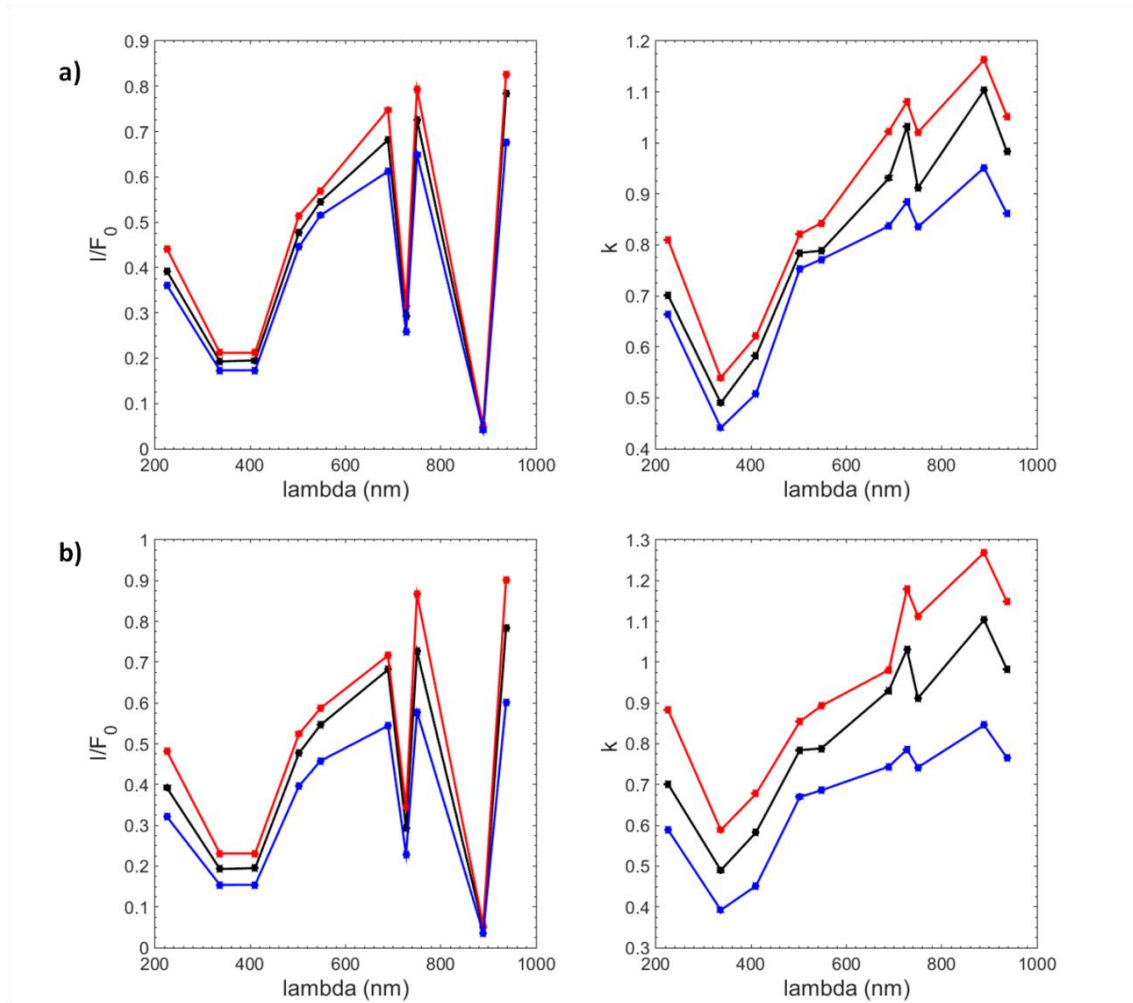

Figure 16: Sensitivity analysis based on changing a given parameter of the tropospheric haze parameterization by 1-σ above (red line) and below (blue line) of the nominal value for the following parameters: a) altitude of the haze base; b) particle peak concentration. The black lines are the values of the best fit data.

## 5. Discussion.

We have found that in the region where ACA system is located, the optical thickness of the stratospheric and tropospheric hazes depends to a large extent on the latitude. This variation is not so pronounced in the lower cloud. As a general rule, we observe that the pressure of the base of the tropospheric haze decreases northwards from north 700 ± 100 mbar to 300 ± 100 mbar.

According to our modelling there are no significant differences between the anticyclones (A) and cyclones (C) either in the stratospheric haze (at pressures levels above tropopause ∼ 60 − 100 mbar) or in the lower cloud (at 1-1.4 bar).

If we compare the reflectivity of the three vortices with the surrounding regions in the range of selected latitudes, we find that in the FQ889N and FQ937N filters the cyclone has low brightness relative to the surroundings, but turns bright in F336W, consistent with low particle density in the tropospheric haze and deeper clouds. A similar behavior has been found in other cyclones (Sánchez-Lavega et al., 2006; Baines et al., 2009). This is in agreement with our result since we found a minimum of optical thickness in the tropospheric haze (~ 10± 2) and a lower concentration of particles (~ 55 ± 10 particles / cm$^3$). On the other hand, anticyclones are brighter than their surroundings in the FQ889N and FQ937N filters and darker in the F336N filter. This situation is associated to two regions with greater optical thickness (~ 25± 2) and a higher concentration of particles (~ 80 ± 10 particles / cm$^3$). Similar observations have been made by Roman et al. (2013) in long-lived cyclones located at 51ºS and by del Río-Gaztelurrutia et al. (2010), who investigated the vertical structure as well as the winds and dynamics.

The particle properties in both regions (effective radius and imaginary refractive index) are similar within our model sensitivity, indicating that for the ACA, no particular difference in microphysics processes and chromophore agents exit. The main difference in the cloud structure between the A and C occurs in the tropospheric haze and affects only to the particle number density and vertical thickness, higher in A than in C. No difference is found between A and C in the base location of this haze, in both cases at ~ 500 mbar. The excess of tropospheric haze particles in A compared to C can be due to a higher ammonia ice condensation in the anticyclones, due to differences in temperature or in the vapor abundance at this level. For Jupiter's anticyclones, dynamical modelling proposes the existence of a cold core above the main cloud deck (Marcus et al., 2012) probably favoring haze formation. This could be also the situation for Saturn anticyclones. On the other hand, the fact that particle density is lower in C than in A could also be due to vertical motions, with subsidence in C at the tropospheric haze, also suggested in a previous work (del Rio-Gaztelurrutia et al., 2010). Note that this is at odds with the behavior that dominates Earth's vortices: subsidence and ample cloud free areas in anticyclones, and cloudy extratropical and tropical cyclones where cloud formation by baroclinic frontal systems and massive moist convection occurs.

## 6. Conclusions.

We report a photometric analysis and radiative transfer modelling of a triple vortex (ACA system) in Saturn's atmosphere using images taken with the HST/WFC3 at a spectral range from 225 nm to 937 nm, including an intermediate and a deep methane absorption band. We retrieve the vertical distribution and properties of the upper cloud and hazes at a region which includes this ACA system, covering a range of latitudes from 55ºN to 69ºN. Below we list the most important conclusions.

- Most atmospheric parameters seem to be zonally homogenous except in the region of the ACA system, where a few of them have significant variations.
- These variations correspond to characteristic parameters of the tropospheric haze, in particular, the particle number density and the base height.
- The optical thickness of the stratospheric haze is quite homogeneous in longitude while increasing with latitude. The optical thickness of the cloud is nearly constant
- The optical thickness of the tropospheric haze increases from $\tau_{trop}$ ~ 28±2 at 55ºN up to $\tau_{trop}$ ~35± 2 at 61ºN , and then decreases down to $\tau_{trop}$ ~10±2 at 69ºN . The greatest variability is found in the range of latitudes of the ACA system. At the anticyclones $\tau_{trop}$ ~20±2, while in the cyclone $\tau_{trop}$ ~ 15±2.
- Both the anticyclones and the cyclone display a base pressure of the tropospheric haze (~550±50 mbar with a greater thickness ~70±2 km = 3H and ~500 ± 50 mbar with a thickness of 50±2 km = 2H), lower than the base pressures of the regions at lower latitudes (750±50 mbar with a thickness 78±2 km = 3.25H to 55ºN ) and higher than the base pressures at higher latitudes (300±50 mbar with a thickness 35±2 km = 1.5H ). H is the atmospheric scale height.
- The maximum particle number density is higher in the anticyclones (~ 78±10 particles/cm$^3$) than in the cyclonic region (~ 50 ± 10 particles/cm$^3$).
- The low values of optical thickness, the concentrations of particles found, as well as the base height of the tropospheric haze in the cyclone suggest that it is a subsidence region.
- The vortices show no significative variations in the scale height, particle size or refractive indices of the haze.

- The properties of the anticyclones and cyclone are compatible with the general picture of upwelling in the former and downwelling in the latter.

**Acknowledgments**

This work was supported by the Spanish project AYA2015-65041-P (MINECO/FEDER, UE) and Grupos Gobierno Vasco IT-765-13.

**References**

Baines, K. H., Delitsky, M. L., Momary, T. W., Brown, R. H., Buratti, B. J., Clark, R. N., and Nicholson, P. D., 2009. Storm clouds on Saturn: Lightning-induced chemistry and associated materials consistent with Cassini/VIMS spectra. Planetary and Space Sci. 57, 1650-1658. doi:10:1016/j.pss.2009.06.025

Del Genio, A.D. , Achterberg, R.K. , Baines, K.H. , Flasar, F.M. , Read, P.L. , Sánchez-Lavega, A. , Showman, A.P. , 2009. Saturn atmospheric structure and dynamics. In: Dougherty, M., Esposito, L., Krimigis, T. (Eds.), Saturn from Cassini-Huygens. Springer-Verlag, pp. 113–159. Chapter 6 .

De Pater, I. and Lissauer, J., 2001. Planetary Science. Cambridge University Press, Cambridge, U.K

del Río-Gaztelurrutia, T., Legarreta, J., Hueso, R., Pérez-Hoyos, S., Sánchez-Lavega, A., 2010. A long-lived cyclone in Saturn's atmosphere: observations and models. Icarus 209, 665–681. doi: 10.1016/j.icarus.2010.04.002 .

del Río-Gaztelurrutia, T., Sánchez-Lavega, A., Antuñano, A., Legarreta,J., García-Melendo, E., Sayanagi, K.M, Hueso, R., Wong, M., Pérez-Hoyos, S., Rojas, J.F., Simon, A.A., de Pater, I., Blalock, J., Barry, T. 2018. A planetary-scale disturbance in a long living three vortex coupled system in Saturn's atmosphere. Icarus 302, 499-513.

Dressel, L., 2019 "Wide Field Camera 3 Instrument Handbook, Version 11.0" (Baltimore: STScI)

Fletcher, L.N., Irwin, P.G.J., Teanby, N.A., Orton, G.S., Parrish, P.D., de Kok, R., Howett, C., Calcutt, S.B., Bowles, N., Taylor, F.W., 2007. Characterising Saturn's vertical temperature structure from Cassini/CIRS.Icarus 189,457–478.

García-Melendo, E., Sánchez-Lavega, A., Hueso, R., 2007. Numerical models of Saturn's long-lived anticyclones. Icarus 191, 665–677. doi: 10.1016/j.icarus.2007.05. 020 .

García-Melendo, E., Pérez-Hoyos, S., Sánchez-Lavega, Hueso,R., 2011. Saturn's zonal wind profile in 2004-2009 from Cassini ISS images and its long-term variability. Icarus 215, 62–74. doi: 10.1016/j.icarus.2011.07.005 .

Hansen, J.E., Travis, L.D., 1974. Light scattering in planetary atmospheres. SpaceSci. Rev. 16, 527–610

Ingersoll, A. P., R. F. Beebe, B. J. Conrath, and G. E. Hunt, 1984. Structure and dynamics of Saturn's atmosphere. In *Saturn* (T. Gehrels and M. S. Matthews,Eds.), pp. 195–238. Univ. of Arizona Press, Tucson.

Ingersoll, A. P., Ewald, S. P.Sayanagi, K. M., & Blalock, J. J. , 2018. Saturn's atmosphere at 1–10 kilometer resolution. *Geophysical Research Letters*, *45*. https://doi.org/10.1029/2018GL079255

Irwin, P.G.J. ,Teanby, N.A., de Kok, R., et al., 2008. The NEMESIS planetary atmosphere radiative transfer and retrieval tool. J. Quant. Spectrosc. Radiat. Transf. 109, 1136–1150.

Irwin, P.G.J. , Fletcher, L.N. , Read, P.L. , et al. , 2015. Spectral analysis of Uranus' 2014 bright storm with VLT/SINFONI. Icarus 264, 72–89 .

Karkoschka, E., Tomasko, M.G., 2005. Saturn's vertical and latitudinal cloud structure 1991–2004 from HST imaging in 30 filters. Icarus 179, 195–221

Lucarini, V., Saarinen, J.J., Peiponen, K.E., Vartiainen, E.M., 2005. Kramers−Kronig Relations in Optical Materials Research.Springer-Verlag Berlin Heidelberg .Germany.

Marcus, P.S., Asay-Davis, X., Wong, M.H., de Pater, I., 2012. Jupiter's new red oval: Dynamics, color, and relationship to jovian climate change. J. Heat Transfer, 135(1), 011007 (9 pages).

Ortiz, J.L , Moreno, M. , Molina, AndA. , 1996. Saturn 1991–1993: clouds and hazes. Icarus 119, 53–66 .

Pérez-Hoyos, S., Sánchez-Lavega, A., French, R.G., Rojas, J.F., 2005. Saturn's cloud structure and temporal evolution from ten years of Hubble space telescope images (1994–2003). Icarus 176, 155–174.

Pérez-Hoyos, S. , Sanz-Requena, J.F. , Sánchez-Lavega, A. , Irwin, P.G.J. , Smith, A. , 2016. Saturn's tropospheric particles phase function and spatial distribution from Cassini ISS 2010–11 observations. Icarus 277, 1–18.

Roman, M.T. , Banfield, D. , Gierasch, P.J. , 2013. Saturn´s cloud structure from Cassini ISS. Icarus 225, 93–110.

Sánchez-Lavega, A., Rojas, J.F., Sada, P.V., 2000. Saturn's zonal winds at cloud level. Icarus 147, 405–420. doi: 10.10 06/icar.20 0 0.6449 .

Sánchez-Lavega, A., Hueso, R., Pérez-Hoyos, S., Rojas, J.F., French, R.G., 2004. Saturn's cloud morphology and zonal winds before the Cassini encounter. Icarus 170, 519–523. doi: 10.1016/j.icarus.20 04.05.0 02

Sánchez-Lavega, A. , Hueso, R. , Pérez-Hoyos, S. , Rojas, J.F. , 2006. A strong vortex in Saturn's south pole. Icarus 184, 524–531.

Sánchez-Lavega A., G. Fisher, L. N. Fletcher, E. Garcia-Melendo, B. Hesman, S. Perez-Hoyos, K. Sayanagi, L. Sromovsky, "*The Great Storm of 2010-2011*", Chapter 13 of the book *Saturn in the 21$^{st}$ Century*, eds. K. H. Baines, F. M. Flasar, N.  Krupp, T. S. Stallard, Cambridge University Press, pp. 377-416 (2019). ISBN 978-1-107-10677-2

Sánchez-Lavega A., E. García-Melendo, J. Legarreta, R. Hueso, T. del Río-Gaztelurrutia, J. F. Sanz-Requena, S. Pérez-Hoyos, A. A. Simon, M. H. Wong, M. Soria, J. M. Gómez-Forrellad, T. Barry, M. Delcroix, K. M. Sayanagi, J. J. Blalock, J. L. Gunnarson, U. Dyudina, S. Ewald, 2019. A complex storm system and a planetary-scale disturbance in Saturn's north polar atmosphere in 2018, Nature Astronomy (submitted).

Sanz-Requena, J.F., Pérez-Hoyos, S., Sánchez-Lavega, A. , et al. , 2012. Cloud structure of Saturn's 2010 storm from ground-based visual imaging. Icarus 219, 142–149.

Sanz-Requena, J.F, Pérez-Hoyos, S., Sánchez-Lavega, A., Antuñano, A., Iwing, P. 2018. Haze and cloud structure of Saturn's North Pole and Hexagon Wave from Cassini/ISS imaging. Icarus .305, 284-300

Sayanagi K., R. West, K. Baines, L. Fletcher, P. Read and U. Dyudina, *"Saturn's Poles: Vortices, Hexagons, and Auroral Aerosols",* Chapter 12 of the book *Saturn in the 21$^{st}$ Century*, eds. K. H. Baines, F. M. Flasar, N.  Krupp, T. S. Stallard, Cambridge University Press, pp. 337-376 (2019). ISBN 978-1-107-10677-2

Smith, B.A, 1981. Encounter with Saturn: Voyager 1 imaging results. Science 212, 163–191. doi: 10.1126/science.212.4491.163.

Smith, B.A, 1982. A new look at the Saturn system: the Voyager 2 images. Science 215, 505–537. doi: 10.1126/science.215.4532.504.

Sromovsky,L.A., Baines, K.H., Fry, P.M. 2013. Saturn's Great Storm of 2010–2011: Evidence for ammonia and water ices from analysis of VIMS spectra. Icarus.226,402-418

Trammell, H.J., Li, L., Jiang, X., Pan, Y., Smith, M.A., Bering, E.A., Hörst, S.M., Vasavada, A .R., Ingersoll, A .P., Janssen, M.A ., West, R.A ., Porco, C.C., Li, C., Si-mon, A .A ., Baines, K.H., 2016. Vortices in Saturn's Northern Hemisphere (2008–2015) observed by Cassini ISS. J. Geophys. Res. Planets 121, 1814–1826. doi: 10. 10 02/2016JE0 05122.

Vasavada, A.R., Hörst, S.M., Kennedy, M.R., Ingersoll, A.P., Porco, C.C., Del Genio, A.D., West, R.A., 2006. Cassinin imaging of Saturn: Southern hemisphere winds and vortices. J. Geophys. Res. 111, E05004.

West,R.A., Baines,K.H., Karkoschka,E., Sánchez-Lavega, A., 2009. Clouds and aerosols in Saturn's atmosphere. In: Saturn from Cassini-Huygens, Dougherty, M.K., Esposito, L.W., Krimigis, S.M.(Eds.).. Springer, Netherlands, pp.161–179.